\documentclass[aps,pre,floatfix,twocolumn,superscriptaddress]{revtex4-2}

\usepackage{mathtools}
\usepackage{xcolor}
\usepackage{amsmath}
\usepackage{amsfonts}
\usepackage{amssymb}
\usepackage{graphicx}
\usepackage{parskip}

\begin{document}

\title{Spatial and particle-particle entanglement in 1D quantum walks of two distinguishable or indistinguishable bosonic particles}

\author{Christopher Mastandrea}
\email{cmastandrea@ucmerced.edu}
\affiliation{Department of Physics, University of California, Merced, CA 95343, USA}

\author{Chih-Chun Chien}
\email{cchien5@ucmerced.edu}
\affiliation{Department of Physics, University of California, Merced, CA 95343, USA}

\begin{abstract}
We present entanglement measures between spatially separated regions and between two distinguishable or indistinguishable particles in one-dimensional two-particle continuous-time quantum walks governed by the Hubbard Hamiltonian. The left-right entanglement checks the entropy of coarse-grained states counting the numbers of particles on the left and right halves of the lattice while the particle-particle entanglement is based on the entropy of the singular values of the time-evolved Fock state. With separable, entangled, and doubly occupied initial states, we examine initial entanglement and the following growth in different entanglement measures.
While the entanglement measures of the indistinguishable cases resemble those of the distinguishable cases when the initial states are comparable, the long-time limits of the entanglement measures are typically non-monotonic as the onsite repulsion increases.
We also discuss possible implications for future research of entanglement in multi-particle quantum dynamics.
\end{abstract}

\maketitle

\section{Introduction}
Quantum walk (QW) is an analogue of the classical random walk created by including the internal degrees of freedom found in the quantum wave function into the description of the walker~\cite{Kempe01072003,QWalkSpringer,Venegas-Andraca2012,KADIAN2021100419, doi:10.1142/S0219749903000383}. The inclusion of these internal states creates markedly different dynamics compared to the classical walk, with the most prominent being a so-called 'ballistic' or quadratic speed up in the spreading time of the walker when compared to the classical walk \cite{PhysRevLett.129.160502}. While the discrete-time QW is determined by a series of alternating operators to shuffle the internal- and real- space distributions, the continuous-time QW and its dynamics are dictated by the time-evolution operator determined by a Hamiltonian. The QW space is typically chosen to be a discreet lattice although other forms of graphs have been investigated \cite{PhysRevA.73.012313, PhysRevA.100.012306}, along with their effect on the correlations within the walk \cite{prerana2026entanglementcapacitycomplexnetworks}. 

Meanwhile, many-body effects have been an important topic intensely studied in condensed matter and atomic, molecular, and optical physics~\cite{coleman2015introduction, RevModPhys.94.041001, Browaeys2020, Eisert2015}. To describe many-body phenomena, Fock-space and many computational techniques have been introduced to investigate strongly correlated systems beyond the single-particle picture~\cite{fehske2007computational, avella2012strongly,Negele2019-ae}. Analogously, multi-particle QWs place a number of individual walkers onto the same space, introducing interactions between particles with previous works investigating many-body topological effects \cite{PhysRevB.108.035126}, networking between quantum computers \cite{IEEE9605308}, and even as a potential model for quantum computations \cite{doi:10.1126/science.1229957}. Therefore, multi-particle QWs offer a complementary route to explore interesting many-body quantum dynamics and offer insights into multi-particle correlations.

On the other hand, the non-classical, no-local correlation known as the entanglement has been under intense investigations due to the lack of a genuine classical analog, making it a wholly quantum effect~\cite{MQM, Carteret1999, eisert2006multiparticleentanglement}. The entanglement generated within many-body systems may give rise to interesting topological ordering of quantum states \cite{doi:10.1126/science.aal3099}. Quantum walks have shown itself to be a promising platform for investigating the generation of entanglement 
within a single walker \cite{PhysRevE.110.064124} or
between multiple walkers \cite{yamagishi2026quantumwalklocalspin, Goyal_2010, PhysRevA.111.052416}. With the recent technological advances ranging from quantum sensing \cite{PhysRevA.97.032329, PhysRevApplied.22.034051, PhysRevLett.114.110506}, to quantum computing \cite{10.1098-rspa.2002.1097}, a great importance has been placed on more precise control and understanding of entanglement effects as a basis for these rapidly advancing quantum technologies.

Experimentally, QWs in both single and multi-particle cases have been realized in various systems such as atomic Bose-Einstein condensates \cite{PhysRevLett.121.070402}, trapped-ions \cite{PhysRevA.65.032310}, or within photonic chips \cite{Zhou2024-qb, Grafe2016-dy, Zhou2024}. When considering the case of particles that can be distinguished from one another, for example two species of atoms or two photons with different frequencies, an internal label can be attached to each particle which is necessary for the distinguishability and is used to formulate the measure of entanglement for these systems, where the von Neumann entropy of the reduced density matrix is typically used \cite{Nielsen_Chuang_2010}. Unlike the case of distinguishable particles, it is not immediately as simple to consider a case such as the entanglement between identical particles due to the inability to differentiate between particles, and attempting to clearly quantify entanglement for systems of indistinguishable particles has been a source of much work over the years \cite{PhysRevLett.91.097902, BENATTI20201, PhysRevA.64.054302, Dalton2017-go, PhysRevX.10.041012}. 

In this work, we focus on two-particle continuous-time QWs that allow for a more systematic and precise investigation into proposed measures of the non-classical correlations between particles for both distinguishable and indistinguishable cases. This restriction to two-particles also greatly reduces the computational resources needed, allowing for investigations into longer time evolutions of the walks. We investigate two different ways of constructing a bipartite system, spatial and particle-particle, and quantify the possible entanglement present from these constructions using entanglement measures taken from working examples with indistinguishable particles in the Fock basis. While the internal label of the distinguishable particles introduces partial trace for quantifying entanglement, we show that some of the formalism can be viewed in a occupation basis form like the indistinguishable particles and thus find entanglement measures that can be generalized to both cases, but some cannot. The long-time dynamics of these entanglement measures are also extracted and compared against various initial states for both cases of particles to differentiate initial-state from genuine dynamical effects.

The rest of the paper is organized as follows. In Section \ref{sec:Theory} we describe the models used to implement the distinguishable and indistinguishable two-particle QWs placed onto a 1D lattice and their time evolution. Section \ref{sec:Measures} introduces and defines the two entanglement measures for investigating spatial and particle-particle entanglement in distinguishable and indistinguishable systems. We show the numerical results obtained from the simulations in Sec. \ref{sec:NumResults} for both cases of particles with a variety of initial states, including separable, entangled, and doubly occupied states. Sec.~\ref{Sec:Implication} discuss some experimental and theoretical implications of entanglement in multi-particle quantum dynamics. We conclude with Section \ref{sec:Conclusion}. The Appendix shows an additional spatial entanglement measure for distinguishable particles and some details of the two-particle QWs.

\section{Theory and simulation} \label{sec:Theory}

\subsection{Quantum walk of two distinguishable particles} \label{sec:DistConstructions}
We begin with the case of two interacting distinguishable walkers on a 1D lattice of size $L$ with open boundary condition. The two walkers are labeled by the spin index $\uparrow,\downarrow$. The quantum walk is governed by a Hubbard Hamiltonian given as
\begin{equation}
    \mathcal{H}_d = -J \sum_{i, \sigma} (c_{i, \sigma}^{\dagger} c_{i+1, \sigma} + H.C) + U\sum_{i} \hat{n}_{i, \downarrow} \hat{n}_{i, \uparrow}
\end{equation}
with $\sigma = (\uparrow, \downarrow)$ the internal degree of freedom of the particles, $J$ and $U>0$ being the hopping coefficient and onsite repulsive coupling constant, and $c_{i}^{\dagger} (c_{i}) $ and $n_i=c_{i}^{\dagger}c_i$ denoting the fermionic creation (annihilation) operator and density operator for a given lattice site $i$. 

We will use $J$ and $t_0=\hbar/J$ as the units of energy and time and set $\hbar\equiv 1$ in the following.

The Fock states for two distinguishable particles are constructed based upon the product states $|\uparrow,l\rangle \otimes|\downarrow, k\rangle$, where $l,k=0,\cdots,L-1$ label the location of the $\downarrow$ $(\uparrow)$ particle in a $L$-site lattice. We number the basis states according to $|k\cdot L+l\rangle$. The basis states are constructed by setting the $\uparrow$ particle at a lattice location, starting with the leftmost site,  and permuting the $\downarrow$ particle through all other lattice locations until all possible states have been generated. We then shift the $\uparrow$ particle to the next lattice site on the right and repeat this process until both particles are placed on the rightmost lattice site at $L$. This construction results in an ordering of states that inherently separates the states corresponding to the $\uparrow$ particle into its left and right sided permutations on the lattice. This creates a total of $L^2$ Fock states for a given lattice size. 
As an example, the basis states for $L=4$ are shown in Table. \ref{table:distinguishableBasisStates}

and from this one can see the way in which the $\downarrow$ particle is permuted around the placement of the $\uparrow$ particle at each lattice site to create each basis state. 

To simulate quantum walks, selected initial states are evolved in time using the time-evolution operator. In our study, we assume the parameters are time independent, so the Hamiltonian can be diagonalized as $\mathcal{H} = UDU^{\dagger}$ with $D$ being a diagonal matrix whose elements are the eigen-energies of the Hamiltonian, and the unitary matrix $U$ containing the eigen-vectors of the Hamiltonian. The time evolution operator $e^{-i\mathcal{H}t}$ can then be written as $e^{-i\mathcal{H}t} = U e^{-iDt} U^{\dagger}$. We evolve a given initial state in time by $|\psi(t+\Delta t)\rangle = U e^{-iD\Delta t} U^{\dagger}|\psi(t)\rangle$. 
Since the formula is exact, we can in principle find the time-evolved state at any moment and extract information for entanglement.

\begin{table}[t]
\parbox{.45\linewidth}{
\centering
\begin{tabular}{l}
$|\uparrow \downarrow, ~0, ~0, ~0\rangle$, \\
$|\uparrow, ~\downarrow, ~0, ~0\rangle$,  \\
$|\uparrow, ~0, ~\downarrow, ~0\rangle$,  \\
$|\uparrow, ~0, ~0, ~\downarrow\rangle$,  \\
$|\downarrow, ~\uparrow, ~0, ~0\rangle$,  \\
$|0, ~\uparrow \downarrow, ~0, ~0\rangle$,  \\
$|0, ~\uparrow, ~\downarrow, ~0\rangle$,  \\
$|0, ~\uparrow, ~0, ~\downarrow\rangle$,  \\
$\vdots$  \\
$|0, 0, 0, ~\uparrow \downarrow\rangle$ \\
\end{tabular}
\caption{Example of the basis-state construction for two distinguishable particles on a lattice of size $L=4$.  \label{table:distinguishableBasisStates}}
}
\hfill
\parbox{.45\linewidth}{
\centering
\begin{tabular}{l}
$|2, ~0, ~0, ~\cdots ~,0\rangle, $ \\
$|1, ~1, ~0, \cdots ~,0\rangle,$ \\
$|1, ~0, ~1, ~0, ~\cdots ~,0\rangle,$ \\
$\vdots$ \\
$|0, \cdots ~,2\rangle.$
\end{tabular}
\caption{Example of the basis-state construction for two indistinguishable bosons on a lattice of size $L=4$.\label{table:indistinguishableBasisStates}}
} 
\end{table}

\subsection{Quantum walk of two indistinguishable particles}\label{sec:IndiConstructions}
In the case of two indistinguishable quantum walkers, we consider the dynamics governed by the 1D Bose-Hubbard model with the Hamiltonian 
\begin{equation}\label{Eq:H_identical}
    \mathcal{H}_i = -J \sum_{i} (a_{i}^{\dagger} a_{i+1} + H.C) + U\sum_{i} \hat{n}_{i} (\hat{n}_{i}-1)
\end{equation}
where $a_{i}^{\dagger} (a_{i}) $ is the bosonic creation (annihilation) operator for a single boson at a lattice site $i$, and $\hat{n}_{i}$ is the number operator. 

Again, the particles are confined on a 1D lattice of size $L$ with open boundary condition.
Since the Hamiltonian shown in Eq.~\eqref{Eq:H_identical} is number conserving, we construct our basis to contain only the Fock states which have a total occupation number of two. This allows for a reduction in the size of the Hilbert space, which leads to a reduction in the memory requirements for the simulations and allows us to move to large lattice sizes that would otherwise be computationally costly.

Explicitly, we construct this reduced basis lexicographically starting with the left-most fully occupied site ($|2, 0, \cdots, 0\rangle$) and permuting through all of the remaining lattice sites. An example of $L=4$ is illustrated in Table~\ref{table:indistinguishableBasisStates}.

To ensure the validity of this restricted model, we used the QuTiP package \cite{qutip5} to generate the same Hamiltonian using the general creation, annihilation operators in the non-restricted space. Using this, we verified that our simulation using the number-conserving basis and the QuTip simulation agree in their final dynamics over a range of lattice sizes up to $L=20$ for the same initial state. Nevertheless, the number-conserving basis allows us to simulate larger systems for longer time periods.

Similar to the distinguishable QW case, we use the exact formula of time-evolution via diagonalization of the Hamiltonian and evolution of the initial states by the corresponding time-evolution operator. Once the time-evolved wavefunction is obtained, we use the entanglement measures described below to extract various kinds of entanglement within the system.

\section{Entanglement Measures} \label{sec:Measures}
Here we describe entanglement measures from partitioning the lattice into left-right components and between the particles. In the following, the $d$ and $i$ subscripts denote the distinguishable and indistinguishable entanglement measures, respectively.

\subsection{Distinguishable-Particle QW}
For two distinguishable particles in a 1D lattice, there are several viable ways to bisect the system. This could be a spatial separation between parts of a well defined lattice, or simply just between the two spin labels.

\subsubsection{Left-Right Entanglement Measure}
\label{sec:DistinguishableLRES}
We first consider splitting the system in real space to a left and right half, and consider the correlations that might be created over the splitting partition.
Following the decomposition laid out in Ref.~\cite{Ident_LR_Entanglement_Book}, the Fock space states can be grouped by the occupation numbers on either side of the partition. These coarse-grained states are given as $|n_{L}, n_{R} \rangle$, where $n_{L} ~ ( n_{R})$ is the total occupation number on the left (right) half of the lattice. For the case of two distinguishable particles (labeled by $\uparrow,\downarrow$) in a 1D lattice, this results in three coarse-grained states, two in which correspond to all of the particles being on either the left or right side of the partition, $|2, 0\rangle, ~|0,2\rangle$, and one in which particles occupy both sides of the partition, $|1,1\rangle$. We will use $j(=0,1,2)$ to label the three possibilities.

We then construct an entanglement measure between the spatial left and right halves of an even-sized lattice by the coarse-grained states. For the time-evolved state $|\psi (t)\rangle = \sum_{n}c_{n}|\psi_n\rangle$, where $n$ denotes the Fock basis states $|k\cdot L+l\rangle$ described in Sec.~\ref{sec:DistConstructions}.
For each of the coarse-grained states in category $j=0,1,2$, we calculate the probabilities $p_{j} = \sum_{n\in j} |c_n|^2$ according to the counting of $n_L,n_R$ for each Fock space configuration $n$. The effective form of the coarse-grained state is then $|\psi(t)\rangle_{cg}=\sum_{j=0}^{2}\sqrt{p_j}|n_L=2-j,n_R=j\rangle$. In this process, only the particle numbers $n_{L,R}$ are of concern, and the spin label does not play a role.
Once the probabilities of the coarse-grained states are obtained, the entanglement measure is defined as
\begin{equation}\label{Eq:ESLR_d}
    ES_{LR,d} = - \sum_{j}p_{j}\ln p_{j},
\end{equation}
where $p_{j}$ are each of the probabilities associated with each category. We remark that the maximal value is $\ln(3)$ according to the definition.

This type of entanglement measure can be straightforwardly generalized to indistinguishable particles, which will be shown shortly. Nevertheless,
since in the case of distinguishable particles the notion of a partial trace over a specific subsystem is well defined, we also investigated another entanglement entropy via partial trace of the density matrix in Appendix \ref{sec:DistinguishableESLRPTraceMethod}. The latter method, however, does not seem to have a counterpart in indistinguishable particles.

\subsubsection{Spin Up-Down Entanglement Measure}
\label{sec:DistinguishableUDES}
From the basis construction shown in Section. \ref{sec:DistConstructions}, we see that it is possible to consider the state to be a product state of the form $|\psi\rangle = \sum_{i,j} c_{i,j} |\uparrow,i\rangle \otimes |\downarrow,j\rangle$. The coefficient matrix $c_{i,j}$ contains correlations between the $|\uparrow,i\rangle$ and $|\downarrow,j\rangle$ states over the whole lattice. If we perform the singular value decomposition (SVD) \cite{strang2006linear, IEE1102314} on $c_{i,j}$, then $c_{i,j} = U\Sigma V^{T}$ where $\Sigma$ is a $L \times L$ diagonal matrix containing the singular values $\lambda_n\ge 0$, $n=1,\cdots,L$.
Performing the SVD on the coefficient matrix is equivalent to finding the coefficients of the Schmidt decomposition~\cite{Nielsen_Chuang_2010}. Explicitly, $|\psi\rangle = \sum_{n} \lambda_{n} |\uparrow,n\rangle \otimes |\downarrow,n\rangle$ where $|\uparrow/\downarrow,n\rangle$ are orthonormal vectors that constitute the Schmidt basis for the $|\uparrow,i\rangle$ and $|\downarrow,j\rangle$ subsystems, respectively.

We may view this decomposed state in another lens in the form of creation operators $a^\dagger_{i,\uparrow}, a^\dagger_{j,\downarrow}$ using the second-quantization. This leads to
$|\psi\rangle = \sum_{i,j}c_{i,j} ~a_{i,\uparrow}^{\dagger}a_{i,\downarrow}^{\dagger}|0\rangle$ with $c_{i,j}$ being the coefficient matrix. Here no symmetrization or anti-symmetrization is imposed, and we assume the operators of different spins commute.  The SVD on the coefficient matrix gives the state $|\psi\rangle = \sum_{n}\lambda_{n} \tilde{a}_{n,\uparrow}^{\dagger}\tilde{a}_{n,\downarrow}^{\dagger}|0\rangle$, where $\tilde{a}_{n,\uparrow}^{\dagger}, \tilde{a}_{n,\downarrow}^{\dagger}$ are the creation operators in the Schmidt basis.
The $\uparrow,\downarrow$ entanglement entropy can then be obtained from
\begin{equation}\label{Eq:UDES_d}
    ES_{\uparrow\downarrow} = -\sum_{n}\lambda_{n}^{2} \ln{\lambda_{n}^{2}}.
\end{equation}
The maximal value is $\ln(L)$ according to the definition. This entanglement measure will naturally generalize to indistinguishable particles, as we will show shortly.

Meanwhile, the above formalism for distinguishable particles is identical to the typical way used to quantify entanglement through entanglement entropy of a partial trace. To see this, we consider the Schmidt-decomposed state given above and construct the full density matrix $\rho = |\psi\rangle\langle\psi| = \sum_{n}\lambda_{n}^{2}|\uparrow, n\rangle |\downarrow, n\rangle \langle\uparrow, n| \langle\downarrow, n|$. 
Taking the partial trace over the $\downarrow$ subsystem, the reduced density matrix is
$\rho_{\uparrow} = Tr_{\downarrow}(\rho) = \sum_{n} \lambda_{n}^{2}|\uparrow, n\rangle \langle \uparrow, n|$, and from this we can see that the entanglement entropy is identical to Eq.~\eqref{Eq:UDES_d}. However, the partial-trace formalism does not apply straightforwardly to identical particles since it is not well-defined when talking about which particle is to be traced out partially. In contrast, the second-quantization method with the SVD of the coefficient-matrix method can be generalized to identical particles.

\subsection{Indistinguishable Particles}
Before proceeding to generalize the aforementioned entanglement measure to identical particles, we caution that the concept of partial trace does not easily apply to identical particles. Instead, we focus on entanglement measures utilizing Fock states. 

\subsubsection{Left-Right Entanglement Measure}
\label{sec:IndistinguishableLRES}
Similar to the case with distinguishable particles, the coarse-grained Fock states $|n_L, n_R\rangle$ also apply to indistinguishable particles. 
Each of these states are constructed by incoherently aggregating the Fock states belonging to each category. For the time-evolved state described by $|\psi(t)\rangle = \sum_{n}c_n|n\rangle$, where $n$ labels the Fock-space basis illustrated in Table~\ref{table:indistinguishableBasisStates},
the effective coarse-grained state is then $|\psi(t)\rangle_{cg}=\sum_{j=0}^{2} \sqrt{p_j}|n_L=2-j, n_R=j\rangle$. The probability is calculated by $p_{j} = \sum_{|n\rangle\in |2-j,j\rangle}|c_n|^2$.
Similar to the distinguishable case, the left-right entanglement measure for indistinguishable particles is then defined as
\begin{equation}\label{Eq:ESLR_i}
    ES_{LR,i} = -\sum_{j=0}^{2} p_{j}\ln p_{j}.
\end{equation}
Again, the maximal value is $\ln(3)$.

\subsubsection{Particle-Particle Entaglement Measure}
\label{sec:InistinguishablePPCM}
Following the discussion given in Refs.~\cite{paskauskas_quantum_2001, Wang_2005}, the Fock state $|\Psi_{B}\rangle$ of two identical bosons can be cast in a second-quantization form as $|\Psi_{b}\rangle = \sum_{i, j=1}^{L} \omega_{i,j} a^\dagger_{i} a^{\dagger}_{j}|0\rangle$ in a 1D lattice of size $L$. 
Here $|0\rangle$ is the vacuum state. The coefficient matrix $\omega_{i,j}$ is a $L \times L$, symmetric matrix which can be decomposed with an appropriate unitary operator $U^{S}$ into the form $\omega_{i,j} = U^{S}Y^{S}(U^{S})^{T}$. From this, the Fock state $|\Psi_{b}\rangle$ can be written in the form $|\Psi_{b} \rangle = \sum_{k=1}^{N}y_{k}^{S}\tilde{b_{k}^{\dagger}}\tilde{b_{k}^{\dagger}}|0\rangle$ where $\tilde{b_{k}^{\dagger}}$ is a bosonic creation operator for mode $k$ in the new Schimdt basis, and $y_{k}^{S}$ is the singular value for mode $k$ from the diagonal matrix $Y^{S}$. This decomposition into a Schimdt basis is shown to exist for any two-particle wave function \cite{PhysRevA.75.062104} and is equivalent to performing the SVD on the coefficient matrix $\omega_{i,j}$ as one can either Schmidt decompose each state, or reshape the states into the coefficient matrix form above and perform the SVD \cite{Nielsen_Chuang_2010}.

This decomposition admits only one single-particle density matrix whose elements $(u, v)$ are given by 
\begin{equation}
\rho_{v,u} = \frac{\langle \Psi_{b}| b^{\dagger}_{u} b_{v} |\Psi_{b}\rangle}{\langle \Psi_{b} | \sum_{u} b^{\dagger}_{u} b_{v} |\Psi_{b}\rangle} = 2(\omega^{\dagger}\omega)_{u,v}.
\end{equation}
The measure that quantifies the entanglement from the uncorrelated state $S_{i} = 0$ to the maximally correlated state $S_{i} = \ln(L)$ is then defined as 
\begin{equation}\label{Eq:Spp_i}
    S_{i} = -\sum_{k=1}^{N} 2|y_{k}^{S}|^2 \ln(2|y_{k}^{S}|^2),
\end{equation}
where $|y_{k}^{S}|$ is the $k$-th singular value from the decomposition. Those factors of $2$ in the above expressions are due to the normalization $a^\dagger_{i} a^{\dagger}_{i}|0\rangle=\sqrt{2}|n_i=2\rangle$.

With this decomposition in the Fock states, the particle-particle entanglement measure is in a form similar to that of the distinguishable particle shown in Sec.~\ref{sec:DistinguishableUDES}. It is important to note, however, that while this decomposition appears to result in a similar form, a partial trace over a given particle remains undefined for indistinguishable particles.

\section{Numerical results} \label{sec:NumResults}
After describing the spatial and particle-particle entanglement measures for both cases of distinguishable and indistinguishable particles, here we show numerical results for each of these measures for both cases and different initial conditions, including separately singly occupied and doubly occupied states.

\subsection{Entanglement of QW of two distinguishable particles}

\subsubsection{Separable initial state}
\begin{figure}[t]
    \centering
    \includegraphics[width=0.8\linewidth]{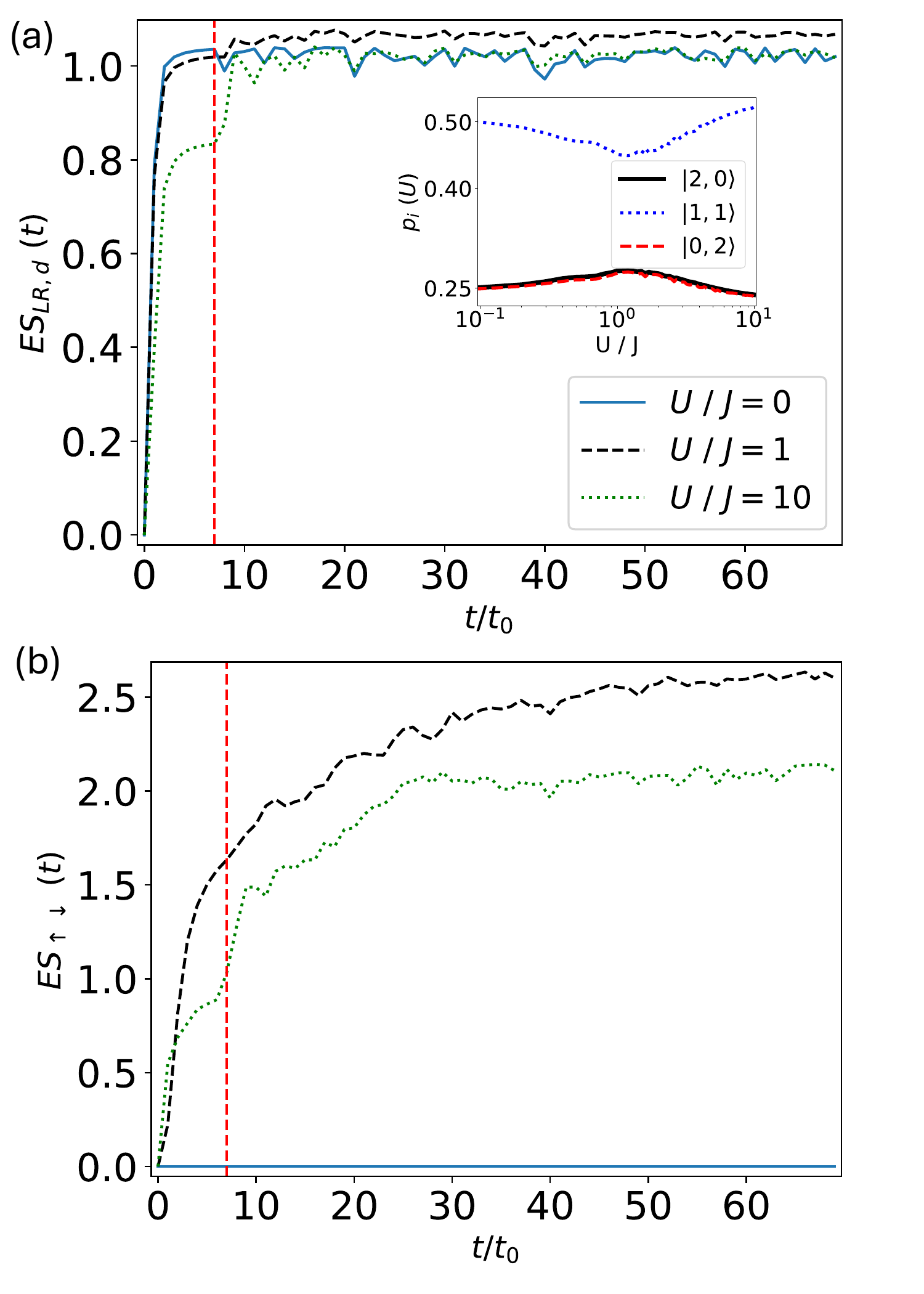}
    \caption{(a) Left-right and (b) $\uparrow, \downarrow$ entanglement measures for two distinguishable particles on a $L=70$ lattice with the separable initial state of  Eq.~\eqref{eq:distinguishableProdInitalState}. The inset of (a) shows the probabilities of the $|n_L,n_R\rangle$ coarse-grained states taken at $t/t_0=20$ for $0 < U/J \leq 10$. In both figures, the red dashed vertical line indicates the time when the particles reach the boundary and reflect off. 
    } \label{fig:DistinguishableFig_nonEnt_doubleFig}
\end{figure}

For two distinguishable particles, we turn our attention first to the case of an un-entangled, separable initial state, placing the particles on adjacent sites in the middle two locations of the lattice. Explicitly,
\begin{equation}
|\psi_{0}\rangle_{sep} = |0, \cdots, 0, \uparrow_{L/2-1}, \downarrow_{L/2}, 0, \cdots, 0\rangle.
\label{eq:distinguishableProdInitalState}
\end{equation}
The spatial left-right entanglement measure given in Eq.~\eqref{Eq:ESLR_d} with the above initial state is shown in Fig. \ref{fig:DistinguishableFig_nonEnt_doubleFig}(a). Looking firstly at the low onsite repulsion curves at $U/J=0$ and $1$, we find that both of these curves follow a similar form in that they both quickly increase within the first few time-steps of the walk as the particles quickly explore both sides of the lattice. For the case of a large onsite repulsion $(U/J=10)$ we see that the entanglement between the two sides of the lattice is reduced compared to the previous cases, however, this is not surprising as the stronger onsite repulsion limits the spreading of the particles within the first time-steps of the walk, restricting the amount of the lattice that can be explored within this time frame.

The vertical red dashed line indicates the time when the particles first reach the boundary of the lattice and get reflected back. Different from discrete-time QWs \cite{QWalkSpringer, KADIAN2021100419, doi:10.1142/S0219749903000383} where a walker moves one lattice site per unit of time, the Hamiltonian evolution of two-particle QWs spreads faster than the time unit $t_0$ indicates because the time-evolution operator has the form of $e^{-i\mathcal{H}t}$. Consequently, the front of propagation gets reflected around $t/t_0\approx 7$ for a $L=70$ lattice for the distinguishable QW with the above initial state. This reflection can be seen in the occupation number plots shown in Appendix~\ref{sec:DynamicsPlots}. We can see the effects of the reflection within the entanglement measure as modulating behavior sets in. 

The inset of Fig. \ref{fig:DistinguishableFig_nonEnt_doubleFig}(a) shows the probabilities for each of the coarse-grained states $|n_L,n_R\rangle$ at $t/t_0=20$ in the steady state as a function of increasing onsite repulsion $U/J$. Here we see a pronounced dip in the probability of the $|1, 1\rangle$ coarse-grained state around $U/J=1$ with the two other coarse-grained states, $|2, 0\rangle$ and $|0, 2\rangle$ both increasing in proportion at the same time. This constriction shifts the probabilities of all three coarse-grained states closer to $1/3$, leading to the maximal left-right entanglement measure around $U/J=1$, as the particles are able to explore doubly occupied states aided by the intermediate onsite repulsion.

In Fig. \ref{fig:DistinguishableFig_nonEnt_doubleFig}(b) we show the $\uparrow, \downarrow$ entanglement measure for two distinguishable particles on a lattice with $L=70$ with the same initial condition for three different values of the onsite repulsion. We find that this entanglement measure grows at a longer time scale compared to the previous case. Although the $U/J=10$ case start exhibiting steady-state behavior, the $U/J=1$ case reaches a steady state a long time later. Moreover, the $\uparrow, \downarrow$ entanglement shows interesting non-monotonic dependence on $U/J$, both before and after the particles reach the lattice boundary and reflect off. Such non-monotonic behavior can be observed for all the entanglement measures in our study, indicating complicated influences on entanglement from competitions between the kinetic and interaction energy scales.

For the $U/J=0$ case in Fig. \ref{fig:DistinguishableFig_nonEnt_doubleFig}(b), $ES_{\uparrow\downarrow}=0$ because without inter-species interactions, the two spin components are completely independent of each other and have no entanglement between them. On the other hand, large $U/J$ tends to suppress doubly occupied states because of their high interaction energy and indirectly reduce the correlation between the two components as well. Therefore, the $\uparrow, \downarrow$ entanglement in the long-time regime is highest for intermediate values of $U/J$, as shown in Fig. \ref{fig:DistinguishableFig_nonEnt_doubleFig}(b).

\begin{figure}[t]
    \centering
    \includegraphics[width=0.8\linewidth]{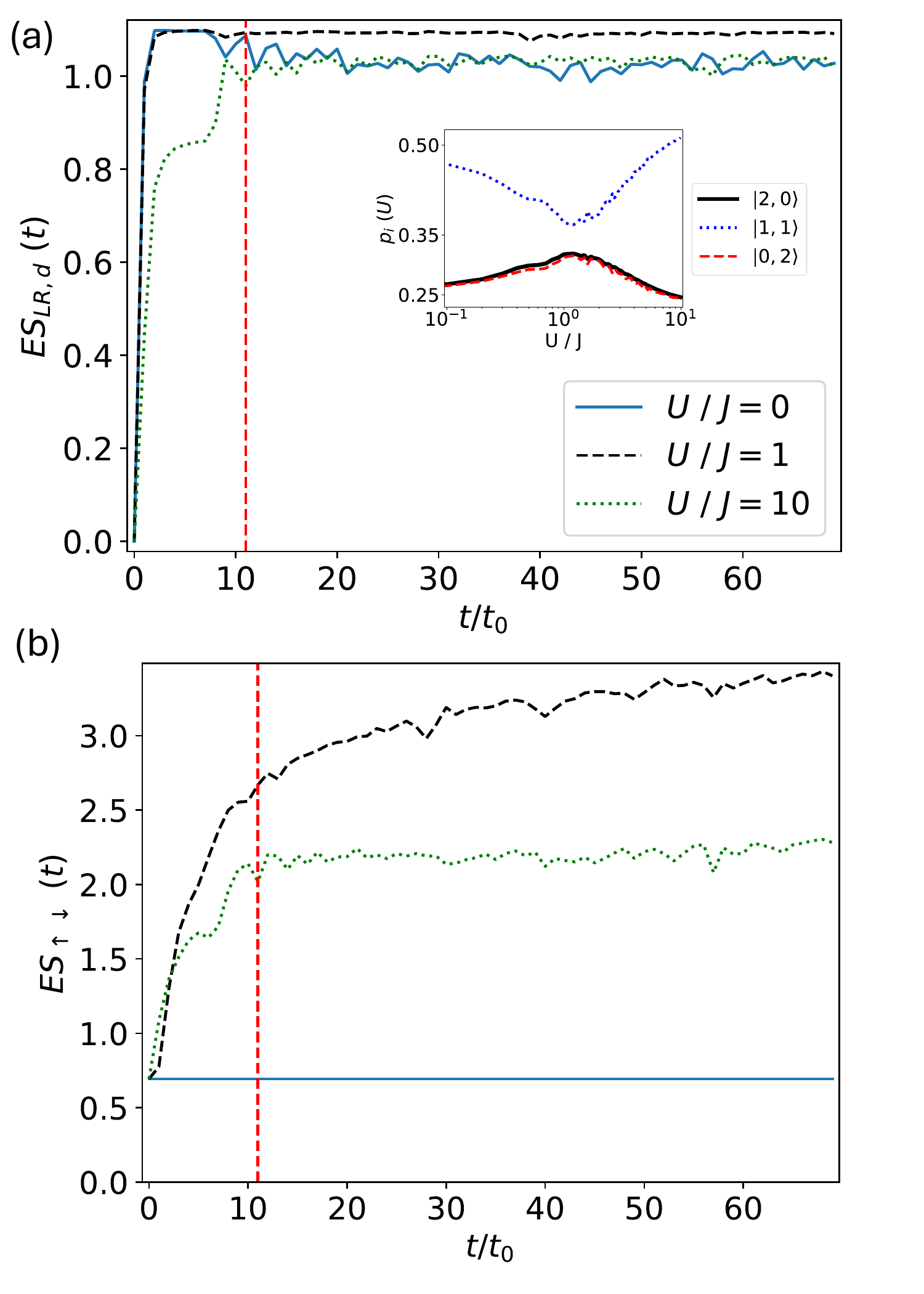}
    \caption{(a) Left-right and (b) $\uparrow, \downarrow$ entanglement measures for two distinguishable particles on a $L=70$ lattice. The initial state is taken as the entangled state shown in  Eq.~\eqref{eq:distinguishableEntInitalState}. The inset in panel (a) shows the probabilities for the corse-grained states for $0 \leq U/t \leq 10$ taken at $t/t_0=20$. For both panels, the red dashed vertical line indicates the time step at which the particles contact the boundary and reflect off.}
    \label{fig:DistinguishableDoubleFig_entangled}
\end{figure}

\subsubsection{Entangled initial state}
Next, we look at the same spatial left-right and $\uparrow, \downarrow$ entanglement measures from an entangled initial state of the two distinguishable particles across the middle two sites of the lattice. Explicitly, 
\begin{eqnarray}
|\psi_{0}\rangle_{ent} &=& \frac{1}{\sqrt{2}} (|0,\cdots,\uparrow_{L/2-1},\downarrow_{L/2},\cdots,0\rangle + \nonumber \\
& &|0,\cdots,\downarrow_{L/2-1},\uparrow_{L/2},\cdots,0\rangle ).
\label{eq:distinguishableEntInitalState}
\end{eqnarray}
We start with the left-right entanglement measure shown in Fig. \ref{fig:DistinguishableDoubleFig_entangled}(a), where one can see a similar trend within the first few steps of the walk as was seen in the separable initial state in Fig.~\ref{fig:DistinguishableFig_nonEnt_doubleFig}(a). In both the cases of low  $(U/J=0, 1)$ and high $(U/J=10)$ onsite repulsion, we see the nearly immediate increase of the entanglement close to its maximal value of $\ln(3) \approx 1.09$ for the low onsite repulsion, with the high repulsion curve sitting below the others. 

When compared to the results from the separable initial state, we do see that the entangled initial state results in the low onsite repulsions getting much closer to the maximum possible value within the first few steps of the walk before the boundary is encountered, although the high onsite repulsion case does not see a similar increase. Past the first reflection from the lattice boundary, we see the entanglement measure falls into a relatively steady value in the long-time limit for all cases of the onsite repulsion. 
We remark that the initial entanglement of the state shown in Eq.~\eqref{eq:distinguishableEntInitalState} is between the two spin components, so there are no significant effects in the left-right entanglement. Indeed, the initial state is of the coarse-grained type $|n_L=1,n_R=1\rangle$ only and thereby has vanishing contribution to the left-right entanglement measure of Eq.~\eqref{Eq:ESLR_d}.

We again show in the inset of Fig.~\ref{fig:DistinguishableDoubleFig_entangled}(a) the probabilities associated with each of the three coarse-grained states at $t/t_0=20$ in the steady state. Here we see that while there is still a clear separation between the $|1, 1\rangle$ and $|2, 0\rangle, |0,2\rangle$ coarse-grained states, at an intermediate value of $U/J$ there is a much more pronounced dip in the likelihood of being in a $|1, 1\rangle$. A similar dip can be seen for the separable initial state in the inset of Fig. \ref{fig:DistinguishableFig_nonEnt_doubleFig}(a), and as this leads to both left-right entanglement measures showcasing similar behaviors between the two initial states, with entangled states reaching a slightly higher value compared to the low- and high- $(U/J)$ cases in the steady state.

In contrast, the $\uparrow, \downarrow$ entanglement measure shown in Fig. \ref{fig:DistinguishableDoubleFig_entangled}(b) exhibits observable effects of the initially entangled state. For all cases of onsite repulsion, the $\uparrow, \downarrow$ entanglement is lifted up to a minimum value of $\approx \ln(2)$ compared to the initially separable state due to the entanglement between the spin degrees of freedom. Past this shift, we find that the entanglement measure for both cases of the initially separable, and initially entangled states produce similar entanglement dynamics within the first few steps of the walk. As the walkers begin to spread within the first few steps of the walk, we find that the increase in the entanglement is much less drastic when compared to the initially separable state. This can be attributed to the same initial lift in the entanglement, effectively jumping past the initial interactions that create the steep, initial increase seen in Fig. \ref{fig:DistinguishableFig_nonEnt_doubleFig}(b). For intermediate values of $U/J$, again it might take a fair amount of time before the $\uparrow, \downarrow$ entanglement settles in to a steady-state value.

\subsubsection{Doubly occupied initial state}
In the previous sections, the initial states for both the separable and entangled initial states were created with particles placed such that the initial lattice states were only singly occupied. Here we start the QW with both particles placed on the same lattice site near the middle. Explicitly,
\begin{equation}
        |\psi_{0}\rangle_{d,d} = |0, \cdots, 0, \uparrow_{L/2}, \downarrow_{L/2}, 0, \cdots0\rangle.
\label{eq:distinguishableRightDoubleOccInitalState}
\end{equation}

\begin{figure}[t]
    \centering
    \includegraphics[width=0.8\linewidth]{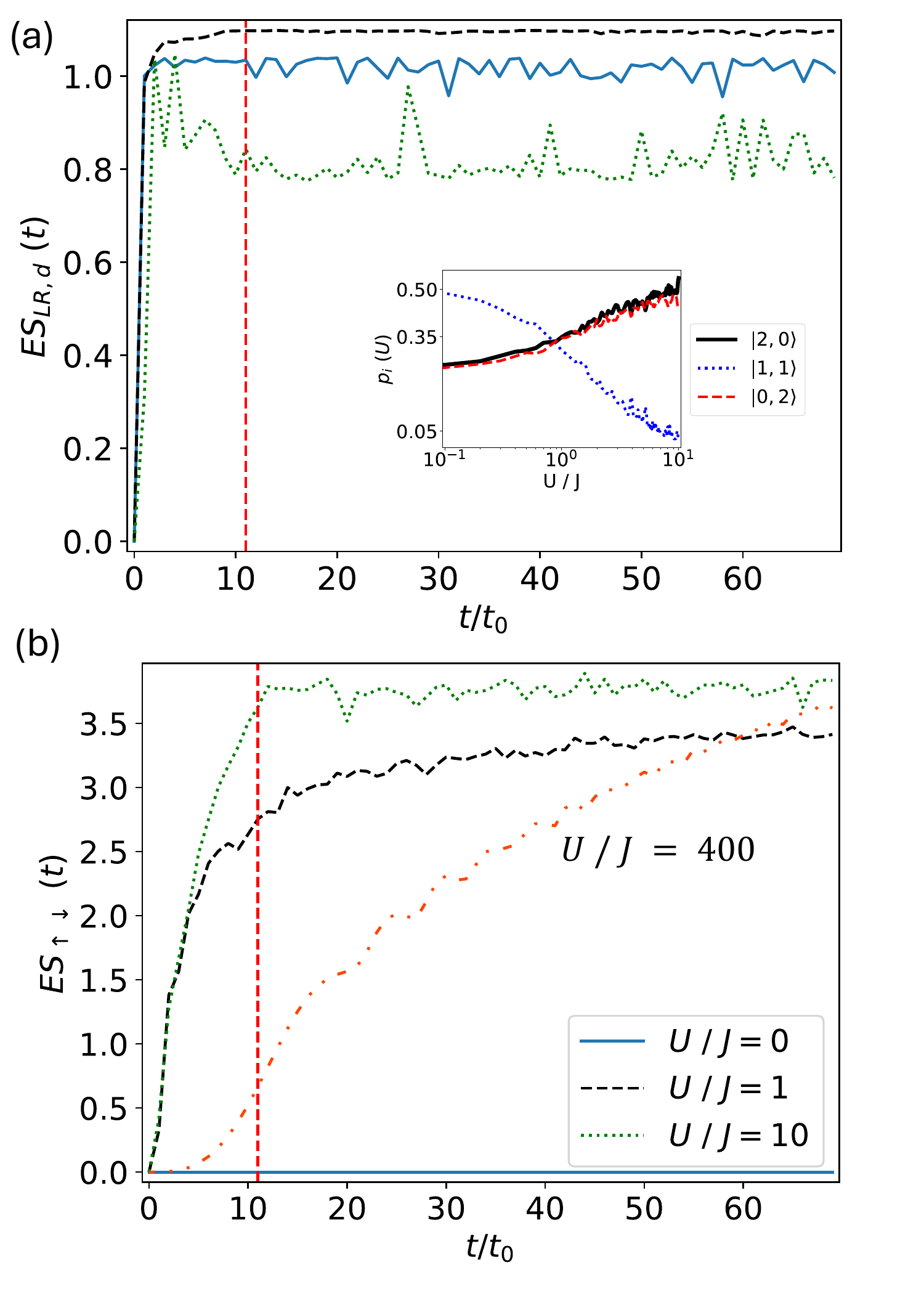}
    \caption{(a) Left-right and (b) $\uparrow, \downarrow$ entanglement measures for two distinguishable particles on a $L=70$ lattice. The initial state is the doubly occupied state shown in Eq. \eqref{eq:distinguishableRightDoubleOccInitalState}. The inset in panel (a) shows the probabilities for the corse-grained states for onsite repulsion values $0 \leq U/J \leq 10$ taken at $t/t_0=20$. The red dashed line indicates the time when the particles first contact the boundary and reflect off. The orange dash-dotted line in panel (b) shows the case of a high onsite repulsion ($U/J=400$).}\label{fig:DistingushableDoubleFig_RightDoubleOccInital}
\end{figure}

The left-right entanglement measure is shown in Fig. \ref{fig:DistingushableDoubleFig_RightDoubleOccInital}(a) for three values of the onsite repulsion $U$. As this initial state places the particles on one side of the lattice partition, we see the same jump in the entanglement entropy associated with the particles moving across the middle of the lattice and beginning to explore the other side. Past this initial state, we find that the change in the onsite repulsion value separates the left-right entanglement curves into three distinct curves in the steady state. 

The inset of Fig. \ref{fig:DistingushableDoubleFig_RightDoubleOccInital}(a) shows the probabilities associated with the three coarse-grained states at time $t/t_0=20$ in the steady state. We see that unlike in the previous cases shown in the inset of Fig. \ref{fig:DistinguishableFig_nonEnt_doubleFig}(a) and Fig. \ref{fig:DistinguishableDoubleFig_entangled}(a) where a clear separation is maintained, the double occupation initial state creates a crossing between the $|1,1\rangle$ and $|2,0\rangle$ ($|0,2\rangle$) coarse-grained states. At an intermediate value of $U/J\sim 1$ when the probabilities cross each other, the maximal value of the left-right entanglement is reached as is seen in the black dashed line for $U/J = 1$ in Fig. \ref{fig:DistingushableDoubleFig_RightDoubleOccInital}(a). Past the crossing point and as the onsite repulsion is increased, the ability for the two walkers to become spatially separated is greatly reduced. Consequently, the left-right entanglement in the steady state tends towards $\ln(2)$ in the large $U/J$ limit.

In the context of the Hubbard model, the so-called atomic limit is reached as $U/J \rightarrow \infty$, and the apparent favor of the multiply-occupied states has been previously investigated and understood \cite{Kapcia_2011, PhysRevB.85.205127, PhysRevLett.104.080401}. Experimentally, doubly occupied states have been called "doublons" and seen with cold-atoms placed in an optical lattice which can be described by the Bose Hubbard model \cite{Winkler2006-or}. 
We remark that here the doubly-occupied states are favored because of the initial state being doubly occupied, and thus energy conservation due to unitary evolution favors doubly occupied states when the interaction energy governed by $U/J$ becomes large.  
Using the left-right entanglement measure of spatial correlations, we can see this same phenomena arise and make its imprint on the entanglement measure with just two particles.

For the $\uparrow, \downarrow$ entanglement measure with the initial double-occupancy state shown in Fig. \ref{fig:DistingushableDoubleFig_RightDoubleOccInital}(b), we see that the overall shape of the curves for each case of the onsite repulsion appear more similar to the entangled initial state shown in Fig. \ref{fig:DistinguishableDoubleFig_entangled}(b), minus the overall shift in the entanglement values since the doubly occupied state for distinguishable particles is not entangled. When $U=0$, the two distinguishable particles are independent of each other, resulting in vanishing entanglement. However, we see that unlike in both the entangled and separable state cases where the higher repulsion tends to limit the entanglement to less than that of the lower repulsion, here we find that the higher repulsion is able to achieve a comparatively greater amount of entanglement compared to the low repulsion case. The high repulsion case also achieves a value much closer to, although still not reaching, the maximum possible value of $\ln(L)\approx 4.24$ for a lattice of size $L=70$.

When $U/J$ increases, the particle-particle entanglement first increases with $U/J$ but then decreases as $U/J$ becomes large. This is understandable since large $U/J$ suppresses quantum dynamics due to the energy-conserving evolution, and the entanglement is generated by dynamics. We show in Fig.~\ref{fig:DistingushableDoubleFig_RightDoubleOccInital}(b) an extra curve of the $\uparrow, \downarrow$ entanglement measure for $U/J=400$ to emphasize the monotonic dependence of the entanglement on $U/J$.

\subsection{QW of two indistinguishable particles}

\subsubsection{Singly occupied, adjacent initial state}
We now look at the entanglement measures defined for the indistinguishable two-particle walk. We start with the adjacent, singly occupied initial state
\begin{equation}
    |\psi_{0}\rangle_{s,i} = |0, \cdots, 0, 1_{L/2-1}, 1_{L/2}, 0, \cdots0\rangle,
\label{eq:indistinguishableAdjInitalState}
\end{equation}
which is a symmetrized bosonic Fock state \cite{Negele2019-ae}. For identical particles, this is the only possible state for distributing two bosons evenly between the two adjacent sites in the middle of the lattice. Therefore, we do not need to differentiate "separable" and "entangled" initial states as we did for the distinguishable cases.

\begin{figure}[t]
    \centering
    \includegraphics[width=0.8\linewidth]{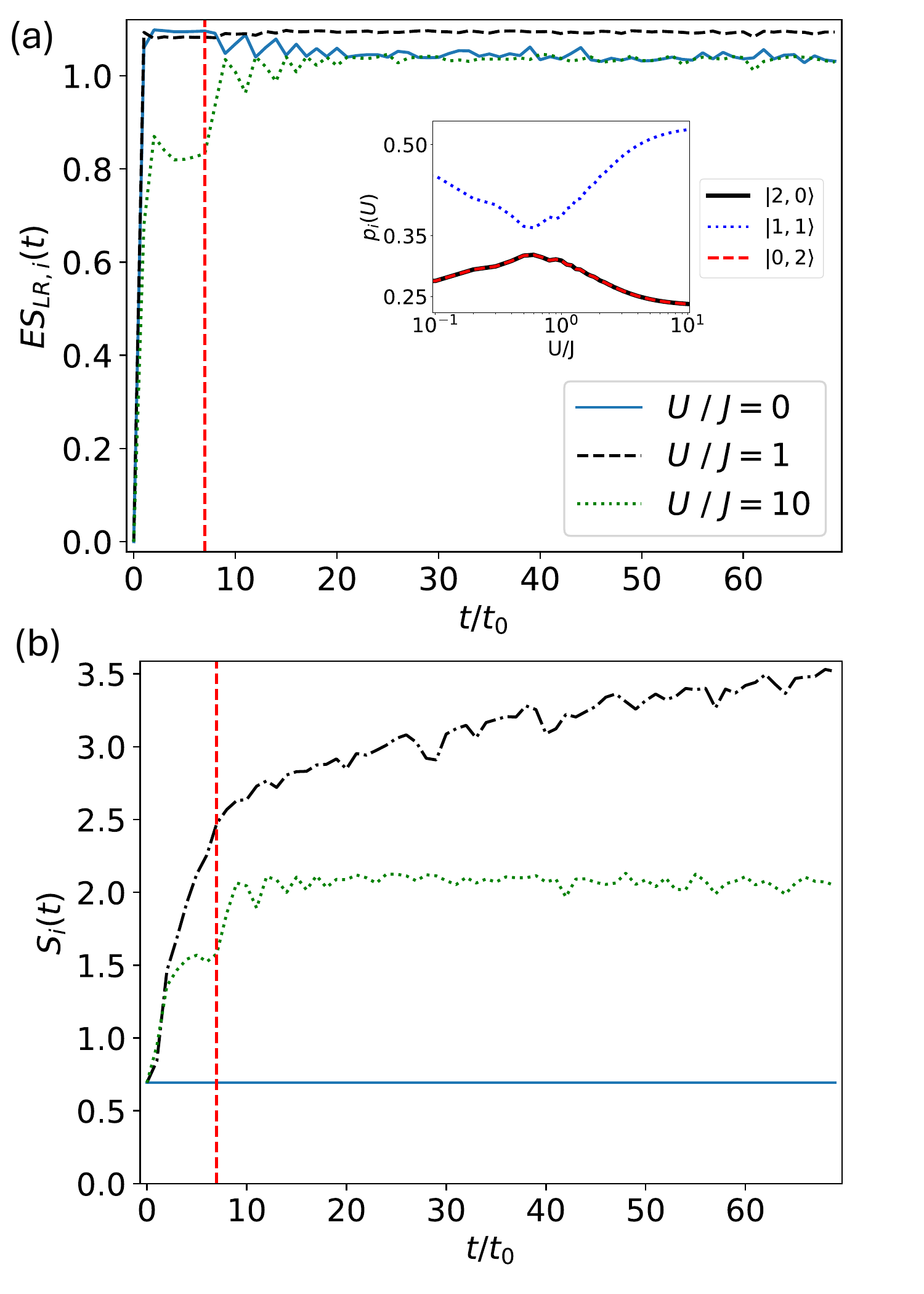}
    \caption{(a) Left-right and (b) particle-particle correlation measures for two indistinguishable particles in a lattice of size $L=70$. The initial state is taken as the adjacent, singly occupied state given in Eq. \eqref{eq:indistinguishableAdjInitalState}. For both panels, the vertical red dashed line indicates the time step where the particles first reach the lattice boundary. The inset in panel (a) shows the probabilities for the corse-grained states for onsite repulsion values $0 \leq U/J \leq 10$ taken at $t/t_0=20$.
    }
    \label{fig:IdenticalFig_midInitial_doubleFig}
\end{figure}

Looking firstly at the left-right entanglement measure, we investigate the same onsite repulsion values as the distinguishable cases and show these results in Fig. \ref{fig:IdenticalFig_midInitial_doubleFig}(a). With this adjacent, singly occupied initial state of indistinguishable particles, we find comparable behavior of the left-right entanglement measure when choosing either separable or entangled initial state for distinguishable particles (shown in Fig. \ref{fig:DistinguishableFig_nonEnt_doubleFig}(a) and Fig. \ref{fig:DistinguishableDoubleFig_entangled}(a), respectively). This is because we effectively treat the distinguishable particles as being in the occupation basis when looking at the spatial entanglement, removing the extra internal degree of freedom. In this respect, indistinguishable particles with only their spatial location being considered exhibit similar spatial entanglement. 

Interestingly, however, we see that the left-right entanglement of indistinguishable particles lie closer to that of the distinguishable case with entangled initial state in Fig.~\ref{fig:DistinguishableDoubleFig_entangled}(a). They both reach closer to the maximal left-right entanglement value of $ES_{LR, max} \approx \ln(3) \approx 1.09$ for the low onsite repulsion cases $(U/J=0,1)$ before the lattice boundary is reached. Moreover, the initial state of Eq.~\eqref{eq:indistinguishableAdjInitalState} is of the type $|n_L=1,n_R=1\rangle$, so the initial left-right entanglement vanishes according to Eq.~\eqref{Eq:ESLR_i}.

In the inset of Fig. \ref{fig:IdenticalFig_midInitial_doubleFig}(a), we can see a similar dip in the likelihood of being in the $|1, 1\rangle$ coarse-grained state at an intermediate value of $U/J$. Just like in the case of the separable and entangled initial states for distinguishable particles, we see that the low- and high- onsite repulsion cases ($U/J = 0$ and $U/J=10$ respectively) converge to similar values in the long-time, with the dip moving the three coarse-grained states closer to an equal probability and allowing the intermediate onsite repulsion to nearly achieve the maximum possible left-right entanglement entropy in the steady state.

We also analyze the particle-particle entanglement measure given in Eq.~\eqref{Eq:Spp_i} with the adjacent, singly occupied initial state and show the results in Fig. \ref{fig:IdenticalFig_midInitial_doubleFig}(b). Similar to the left-right entanglement measure, we see that the particle-particle entanglement measure has a form closer to the entangled initial state for distinguishable particles seen in Fig. \ref{fig:DistinguishableDoubleFig_entangled}(b), even with the same $\ln(2)$ shift. For two indistinguishable particles, what looks like an initially separable state with adjacent, singly occupied sites shows an initial entanglement between the particles due to the symmetrization of bosonic Fock states. The initial state of Eq.~\eqref{eq:indistinguishableAdjInitalState} has a coefficient matrix $\omega_{ij}$ with only two non-zero values at the off-diagonal locations $(L/2 -1, L/2)$ and $(L/2, L/2-1)$.

The $2\times2$ block of the matrix which contains the non-zero values  looks much like the Pauli matrix $\sigma_{x}$ (with an additional factor of $1/2$ from the bosonic symmetrization).

Thus, when performing the SVD on $\omega_{ij}$, only two non-vanishing singular values of $1/2$ will be obtained, making the initial particle-particle correlations $S_{i}(t=0) = -2(\frac{1}{2}\ln(\frac{1}{2})) = \ln(2)$ according to Eq.~\eqref{Eq:Spp_i} for this initial state, creating a shift in entanglement for any onsite repulsion value.

We remark that Fig.~\ref{fig:IdenticalFig_midInitial_doubleFig} resembles Fig.~\ref{fig:DistinguishableDoubleFig_entangled}. Therefore, two distinguishable walkers with an entangled initial state exhibit similar entanglement in real space and between the particles as that of two indistinguishable with adjacent, singly occupied initial entangled state. However, the case with distinguishable particle and separable initial state shown in Fig.~\ref{fig:DistinguishableFig_nonEnt_doubleFig} can be differentiated by the entanglement between the particles.

\subsubsection{Doubly occupied initial state}
We also analyze the entanglement measures with a doubly occupied initial state for indistinguishable particles given by
\begin{equation}
        |\psi_{0}\rangle_{d,i} = |0, \cdots, 0, 2_{L/2}, 0, \cdots0\rangle.
\label{eq:indistinguishableRightDoubleOccInitalState}
\end{equation}

\begin{figure}[t]
    \centering
    \includegraphics[width=0.8\linewidth]{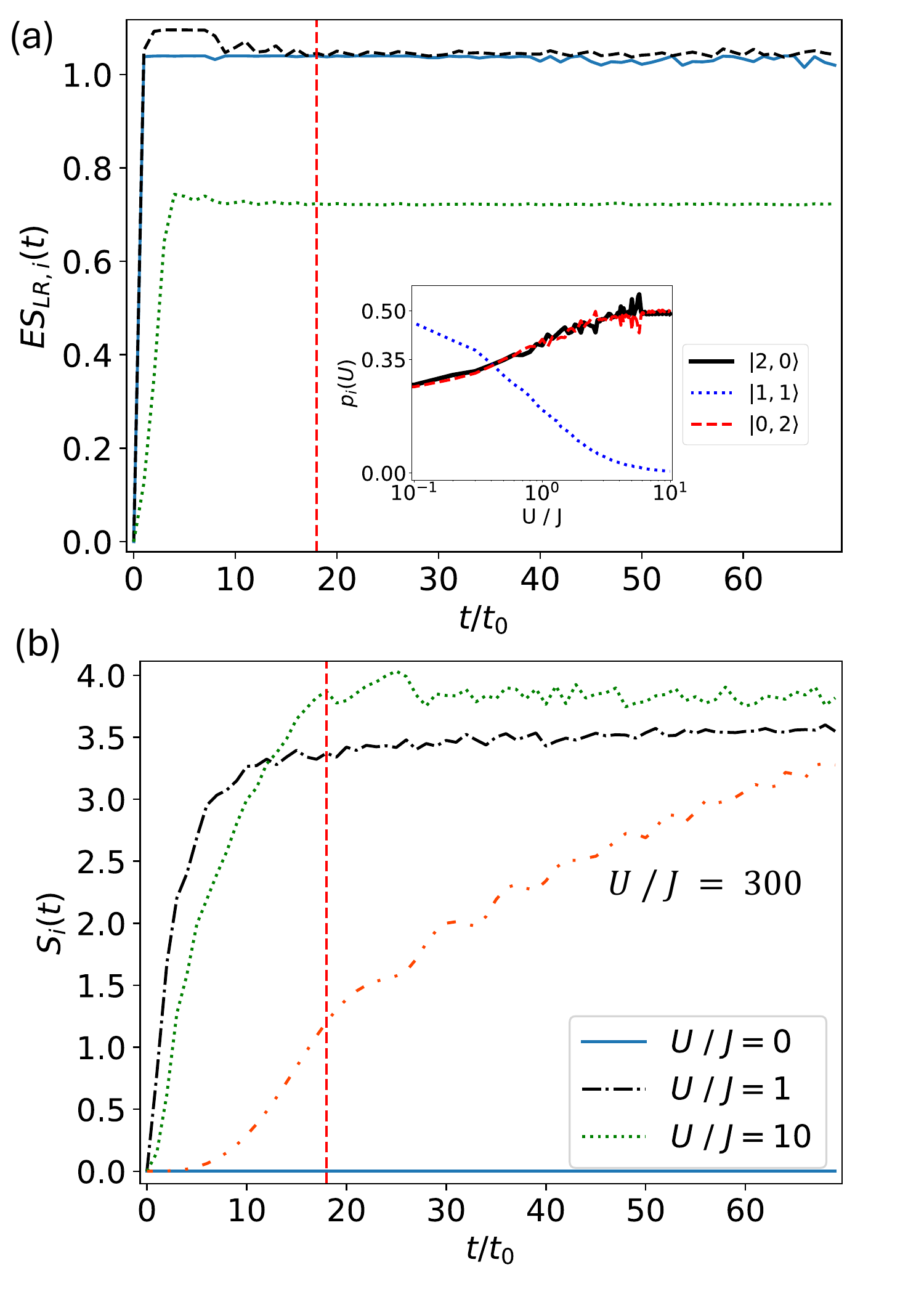}
    \caption{(a) Left-right and (b) particle-particle correlation measures for two indistinguishable particles in a lattice of size $L=70$. The initial state is taken as the doubly occupied state of  Eq.~\eqref{eq:indistinguishableRightDoubleOccInitalState}. The red dashed line indicates the time at which the particles first reach the boundary and reflect off. The inset in panel (a) shows the probabilities for the corse-grained states for onsite repulsion values $0 \leq U/t \leq 10$ taken at $t/t_0=20$. The orange dash-dotted line in panel (b) shows the case of a high onsite repulsion ($U/J=300$).
    }  \label{fig:indistingushableFig_RightDoubleOccInital_doubleFig}
\end{figure}

Fig. \ref{fig:indistingushableFig_RightDoubleOccInital_doubleFig}(a) shows the left-right entanglement entropy for different values of the onsite repulsion. For low values of onsite repulsion $(U/J=0, 1)$, we see that the left-right entanglement entropy is able to reach or come close to the maximum possible value within the first few steps of the walk. More interestingly, however, we see that a large repulsion $(U/J=10)$ greatly reduces the maximum steady-state value of the left-right entanglement. 

The inset of Fig. \ref{fig:indistingushableFig_RightDoubleOccInital_doubleFig}(a) shows the probabilities for each of the three coarse-grained states ($|n_L,n_R\rangle=|2,0\rangle, |1,1\rangle, |0,2\rangle$) as a function of the onsite repulsion at $t/t_0=20$ in the stead state. We can see that an apparent crossing occurs around $U/J\approx 0.4$ where the likelihood of the particles being split across the lattice partition is greatly reduced, with a much greater chance of them remaining in a doubly occupied state in the long-time limit of the walk. The reason that the $|2,0\rangle,|0,2\rangle$ coarse-grained states are favored when $U/t\gg 1$ is because of the unitary time evolution that conserves energy. Since the initial state is a doubly occupied state with higher potential energy due to the large $U/t$, its dynamics favors the two coarse-grained states with double occupancy. Consequently. only two coarse-grained states are at play in the steady state, and the resulting left-right entanglement measure drops to $\ln(2)$.
Interestingly, influences on unitary quantum dynamics due to double occupancy is not limited to distinguishable and indistinguishable bosonic particles. A previous work~\cite{Chien_2013} has shown that doubly occupied states affects quantum transport of identical fermions with tunable interactions as well.

As the onsite repulsion is increased, we find that for the lower repulsion value shown as the dashed black line ($U/J=1$), it can be seen that the spatial left-right entanglement reaches a steady state value within $t/t_0=5$, while the larger repulsion value ($U/J=10$) can create more entanglement between the left and right halves of the lattice but requires almost three times the amount of time ($t/t_0\approx 18$) before it reaches a steady-state value. However, the steady-state left-right entanglement is not monotonic with $U/J$ since it slightly increases from $U/J=0$ to $U/J=1$ and then decreases for $U/J=10$.

We now turn to the particle-particle entanglement measure of two indistinguishable particles shown in Fig.~\ref{fig:indistingushableFig_RightDoubleOccInital_doubleFig}(b) 
when staring with a double occupancy initial state in Eq.~\eqref{eq:indistinguishableRightDoubleOccInitalState}. The initial particle-particle entanglement is zero because the doubly occupied initial state already takes the second quantization form $(1/\sqrt{2})a^\dagger_{L/2}a^\dagger_{L/2}|0\rangle$ with only one non-vanishing singular value, thereby causing Eq.~\eqref{Eq:Spp_i} to produce zero particle-particle entanglement.

With the initial entanglement understood, we now look at the dynamics of the correlation measure as the system is evolved in time. For cases of a non-zero onsite repulsion $(U/J = 1, 10)$, we see that the entanglement between the two particles is able to attain much higher values when compared to the same curves for the adjacent, singly occupied initial state shown in Fig. \ref{fig:IdenticalFig_midInitial_doubleFig}(b). We note that especially in the case of $U/J = 1$, this measure does not appear to saturate within the monitored evolution time, although hints of saturation might be seen for the case of $U/J=10$.

The particle-particle entanglement measure for two indistinguishable particles with the initial doubly-occupied state of Eq.~\eqref{eq:indistinguishableRightDoubleOccInitalState} and $U=0$ remains zero during time evolution, implying the system remains in a particular Fock state even though the particles are hopping in the lattice. This can be understood by first analyzing the case of a lattice with periodic boundary condition for simplicity.
For a periodic lattice with $L$ sites and two identical particles with $U=0$, the Hamiltonian of Eq.~\eqref{Eq:H_identical} only has the hopping terms.
 
The lattice Fourier transform gives the Hamiltonian $\mathcal{H}_i = -\sum_{q}\varepsilon_{q}c^\dagger_{q}c_{q}$. Here the creation operator on site $l$ is written as $c_{l}^{\dagger} = \frac{1}{\sqrt{L}}\sum_{k}c^{\dagger}_{k}e^{2\pi ikl/L}$, and $\varepsilon_{q}$ is the single-particle energy dispersion. For the initial condition with both particles on the same lattice site, we use translation symmetry to place them at site $l=0$ for simplicity. Therefore, $|\psi_{0}\rangle = c_{0}^{\dagger}c_{0}^{\dagger} |0\rangle = \frac{1}{L}\sum_{k}\sum_{k^{'}}c^{\dagger}_{k}c^{\dagger}_{k^{'}}|0\rangle$, where $|0\rangle$ is the vacuum state.
Applying the time-evolution operator $\exp(-i\mathcal{H}_it)$ on the initial state gives the evolved state $|\psi(t)\rangle = e^{t\sum_{q}\varepsilon_{q}c^\dagger_{q}c_{q}}\frac{1}{L}\sum_{k}\sum_{k^{'}}c^{\dagger}_{k}c^{\dagger}_{k^{'}}|0\rangle$. The density operator $n_q=c^\dagger_{q}c_{q}$ extracts the number of particles in the two Fourier modes $k, k^{'}$. Consequently, $|\psi(t)\rangle = \frac{1}{L}\sum_{k, k^{'}}e^{t\varepsilon_{k}}e^{t\varepsilon_{k^{'}}}c^{\dagger}_{k}c^{\dagger}_{k^{'}}|0\rangle$, resulting in the coefficient matrix $\beta_{k, k^{'}} =\frac{1}{L}\sum_{k, k^{'}}e^{t\varepsilon_{k}}e^{t\varepsilon_{k^{'}}}$.
If we perform the SVD on the coefficient matrix while scaling the orthogonal matrix  $V$ by the exponential factors in the coefficient matrix, the scaled coefficient matrix contains $1$ for every element. The resulting singular values have only one non-vanishing entity $\lambda_{1} = L$ with the eigenvector $(1,1,\cdots)^T$, and all the other singular values vanish. Hence, the evolved state effectively remains in one doubly occupied state over the course of the QW, and no particle-particle entanglement could be generated according to Eq.~\eqref{Eq:Spp_i} since the state takes the form $\hat{c}^\dagger_1\hat{c}^\dagger_1|0\rangle$, where $\hat{c}^\dagger_1$ is the non-vanishing transformed state after the SVD. We caution that the time-evolved state is doubly occupied neither in real space nor in momentum space. For open boundary condition used in this work, the argument goes through if the lattice Fourier transform is replaced by the lattice sine transform. It is thus interesting to see that no particle-particle entanglement can be generated if $U=0$, regardless of the initial states studied above.

For intermediate values of $U/J$, the particle-particle entanglement becomes finite. As explained before, however, large $U/J$ suppresses quantum dynamics and leads to lower particle-particle entanglement. To emphasize the non-monotonic dependence of the long-time entanglement on $U/J$, we show an extra curve of the particle-particle entanglement for $U/J=300$ in Fig.~\ref{fig:indistingushableFig_RightDoubleOccInital_doubleFig}(b), which falls below the other cases ($U/J=1,10$) as expected.

Looking more generally at both cases of particles considered (distinguishable and indistinguishable) and all of their respective initial states investigated in this work, we can begin to see comparable behavior between some of these cases. For example, when looking only at the left-right entanglement, we find that the doubly occupied initial states for both cases of particles will show an effective transition of the probabilities of the coarse-grained states at an intermediate value of the onsite repulsion. Interestingly, the choice of particles does not change the overall trend from this transition, that being the reduction in the left-right entanglement measure as $U/J$ becomes large, though more fluctuations past this transition can be seen in the left-right entanglement measure for the distinguishable particles compared to the indistinguishable case.

A comparison between Fig.~\ref{fig:DistingushableDoubleFig_RightDoubleOccInital} and Fig.~\ref{fig:indistingushableFig_RightDoubleOccInital_doubleFig} thus shows the resemblance of the entanglement measures of distinguishable and indistinguishable QWs with doubly-occupied initial states. Moreover, the similarities support the generalizations of the entanglement measures based on the coarse-grained states and Fock states from distinguishable to indistinguishable cases. Those generalizations of entanglement measures thus allows us to focus on the mechanisms for entanglement generation in multi-particle quantum dynamics.

\section{Implications}\label{Sec:Implication}
QWs have seen a variety of different experimental realizations such as in analogue photonic systems, atomic Bose-Einstein condensates, and even including digital simulations on current noisy intermediate-scale quantum (NISQ) computers \cite{Acasiete2020, PhysRevA.103.022408, PhysRevResearch.5.033089}. Attempting to experimentally reconstruct the state of a quantum system has long been of interest. In recent years, various works in quantum state tomography have progressed that allow for efficient reconstruction of a quantum state and its content \cite{Cramer2010-bx, Karski2009-ym, PhysRevLett.74.884, PhysRevLett.133.160801}. In particular, Ref. \cite{PhysRevLett.111.020401} demonstrates a method for efficiently reconstructing the density matrix of a 1D, many-body system requiring only local measurements of the expectation values of a minimal set of observables that can be used to fully describe the state. 
Once the quantum state or its density matrix is obtained, the aforementioned entanglement measures can be applied to quantify entanglement in real space or between particles and characterize the quantum nature of the system of interest.

We remark that the choice of distinguishable particles and indistinguishable bosons in this work are motivated by experimental implementations of photons and atoms mentioned above. Nevertheless, the left-right and particle-particle entanglement measures can be straightforwardly generalized to identical fermions. Of course, the Pauli exclusion principle enforced by the anti-commutation relations of fermionic operators will lead to different behavior in the entanglement measures. Studies of quantum dynamics of many electrons or fermionic atoms may be analyzed in a similar fashion to extract information about entanglement \cite{PhysRevA.95.062320, LoFranco2016}.
Moreover, recent experiments working with free electrons have implemented a two-particle QW, performing a two-electron state tomography method to reconstruct the combined state of the particles, further investigating the entanglement between free electrons \cite{Tziperman2026}. Therefore, entanglement in multi-particle quantum systems attracts broad interest and continues to advance our understanding of quantum dynamics.

As discussed before, other works on multi-particle QWs have investigated the correlations generated between the interactions with walkers. However, some only consider distinguishable particles with internal spin labels~\cite{Giri2021}. 
Although Ref. \cite{yamagishi2026quantumwalklocalspin} use symmetrization and antisymmetrization of wave-functions to investigate QWs of identical particles, it implements
entanglement measures such as the negativity that while easily calculable, requires the partial trace over a sub-system and thus encounter issues when attempting to directly apply them to indistinguishable particles, particularly for particle-particle entanglement. The alternative left-right entanglement measure based on partial trace in Appendix~\ref{sec:DistinguishableESLRPTraceMethod} also cannot be generalized to indistinguishable particles straightforwardly. It thus remains a challenging but rewarding task to define and evaluate more entanglement measures for multi-particle quantum dynamics involving distinguishable as well as indistinguishable particles.

\section{Conclusion} \label{sec:Conclusion}
We have shown two entanglement measures, based on spatial and particle correlations of the Fock states, applicable to distinguishable as well as indistinguishable bosonic particles in continuous-time QWs. Through the simulations of the exact quantum dynamics and analyses of the time-evolved wavefunctions with various singly- and doubly-occupied initial states, we show the indistinguishable cases resemble the distinguishable cases with comparable initial states, thereby illustrating the generality of the multi-particle entanglement measures. Moreover, the variety of entanglement in QWs unveil intricate internal correlations among different subsystems during quantum dynamics. With the rapid advancement in realizations and manipulations of many-body quantum systems, our work offers inspiring aspects of entanglement and its applications in QWs.

\begin{acknowledgments}
This work was supported by the DOE (Grant No. DE-SC0025809). This research was conducted using Pinnacles (NSF MRI, No. 2019144) at the Cyberinfrastructure and Research Technologies (CIRT) at University of California, Merced.
\end{acknowledgments}

\appendix

\section{Another left-right entanglement measure for distinguishable particles} \label{sec:DistinguishableESLRPTraceMethod}
When looking at measuring the spatial left-right entanglement for distinguishable particles, it is possible to formulate another entanglement measure by taking advantage of the spatial splitting that is inherit in our basis construction.
As mentioned when discussing the basis states for two distinguishable particles, the chosen ordering automatically arranges the left and right occupation states when considering only the $\uparrow$ particle.
In terms of the basis $|k\cdot L+l\rangle$ illustrated in Table~\ref{table:distinguishableBasisStates}, a state may be viewed as 
$|\psi_{tot}\rangle = \begin{pmatrix} |\psi_{L, \uparrow}\rangle  \\ |\psi_{R, \uparrow}\rangle \end{pmatrix}$ since the $\downarrow$ particle arranges its location repeatedly while the $\uparrow$ particle is arranged from the left to the right. The density matrix can then be constructed as $\rho_{tot} = |\psi_{tot}\rangle \langle\psi_{tot}|$, which may be though of as being split into four distinct blocks of size $L^{4}/4$, following the separation of the components of the state into left and right parts. These blocks are laid out as 
\begin{math}
\rho_{tot} = 
\begin{pmatrix}
    \rho_{LL} & \rho_{LR} \\
    \rho_{RL} & \rho_{RR}
\end{pmatrix}.
\end{math}
We can extract a reduced density matrix by taking the trace of each of these blocks,  giving 
\begin{math}
\rho_{redu} = 
\begin{pmatrix}
    Tr(\rho_{LL}) & Tr(\rho_{LR}) \\
    Tr(\rho_{RL}) & Tr(\rho_{RR})
\end{pmatrix}
\end{math}
which is now just a $2 \times 2$ matrix.

\begin{figure}[t]
    \centering
    \includegraphics[width=0.8\linewidth]{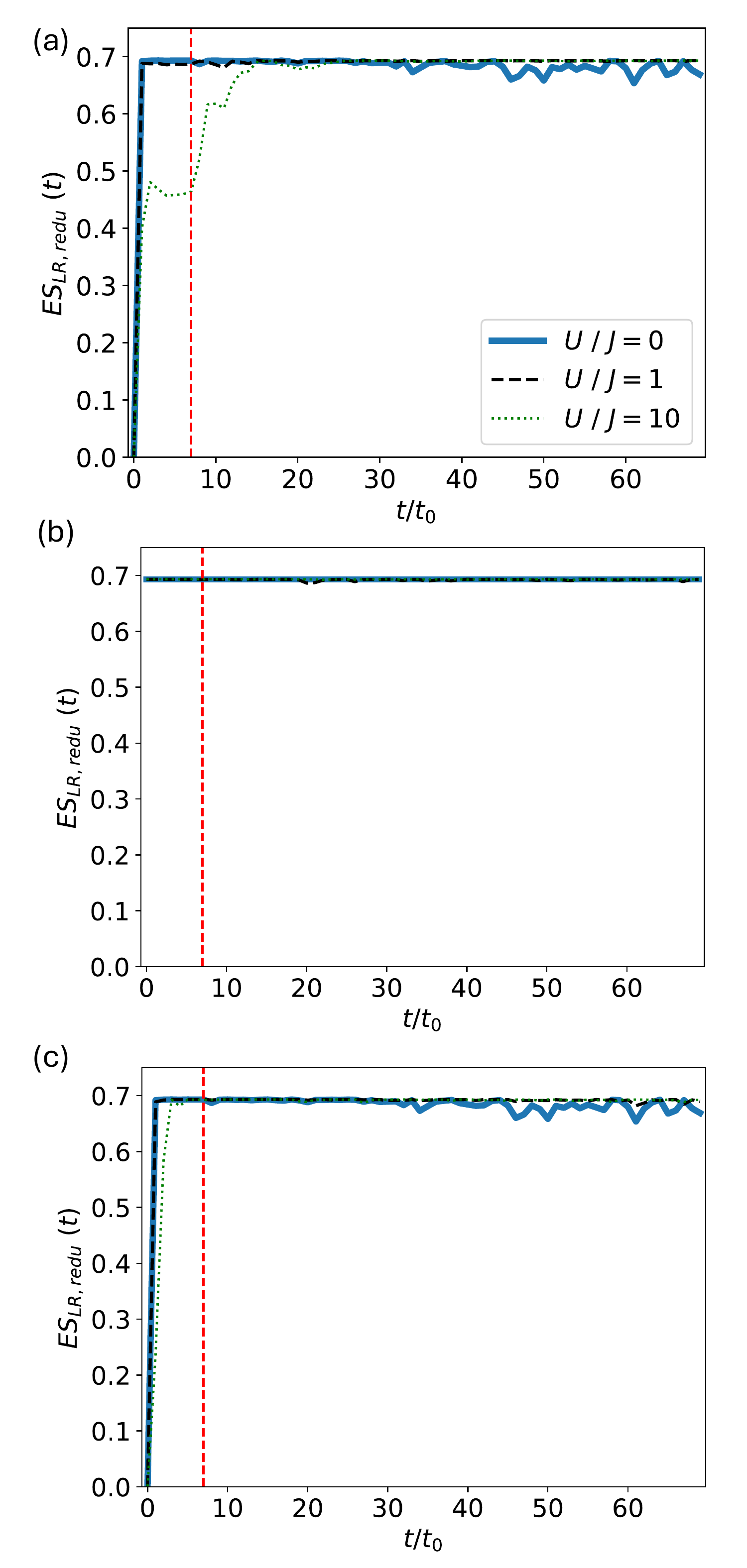}
    \caption{Alternative left-right entanglement entropy of Eq. \eqref{eq:altLeftRightESDefinitionDistinguishable} for two distinguishable particles in a lattice of size $L=70$. The initial state is taken as (a) separable, (b) entangled, and (c) double occupancy initial state given in Eq. \eqref{eq:distinguishableProdInitalState}, Eq. \eqref{eq:distinguishableEntInitalState}, and Eq. \eqref{eq:distinguishableRightDoubleOccInitalState} respectively. The red dashed vertical line indicates the time when the particles reach the boundary and reflect off.
     }
    \label{fig:DistinguishableESLRReduDoubleFig}
\end{figure}

The reduced density matrix contains the correlations between the left and right states of the $\uparrow$ particle since other degrees of freedom have been traced out. Thus, we can compute this entanglement measure in the typical fashion with the entanglement entropy \cite{Nielsen_Chuang_2010}
$ES_{LR,redu} = -Tr(\rho_{redu}\ln{\rho_{redu}})$. This can be simplified in terms of the two eigenvalues $\lambda_{1,2}$ of the reduced density matrix via
\begin{equation}
    ES_{LR,redu} = -\lambda_1 \ln\lambda_1 - \lambda_2 \ln\lambda_2.
    \label{eq:altLeftRightESDefinitionDistinguishable}
\end{equation}
The maximal value is $\ln(2)$ according to the definition.

This spatial left-right entanglement measure for a separable initial state of Eq.~\eqref{eq:distinguishableProdInitalState} is plotted in Fig. \ref{fig:DistinguishableESLRReduDoubleFig}(a). We see that for comparatively small values of the onsite repulsion $(U/J<1)$ the entanglement measure shows similar dynamics to each other with both reaching the maximal value of $ES_{LR,redu} \approx \ln(2)$ shortly after the start of the walk. This quick jump and constant value can be seen as coming from each particle having a roughly equal chance of it moving to either side of the lattice, leading to a quick mixing between the particles on either side of the lattice. 
For $U/J\geq10$, we see that while the left-right entanglement $ES_{LR,redu}$ reaches the maximum value in the long-time limit, it does not follow the rapid increase that the other curves show. There is a dip in the entanglement measure soon after the start of the walk. Shortly after this dip we see the main jump that brings it closer to the maximal value occurs just after the wavefunction first interacts with the boundary of the lattice.

\begin{figure}[t]
    \centering
    \includegraphics[width=0.8\linewidth]{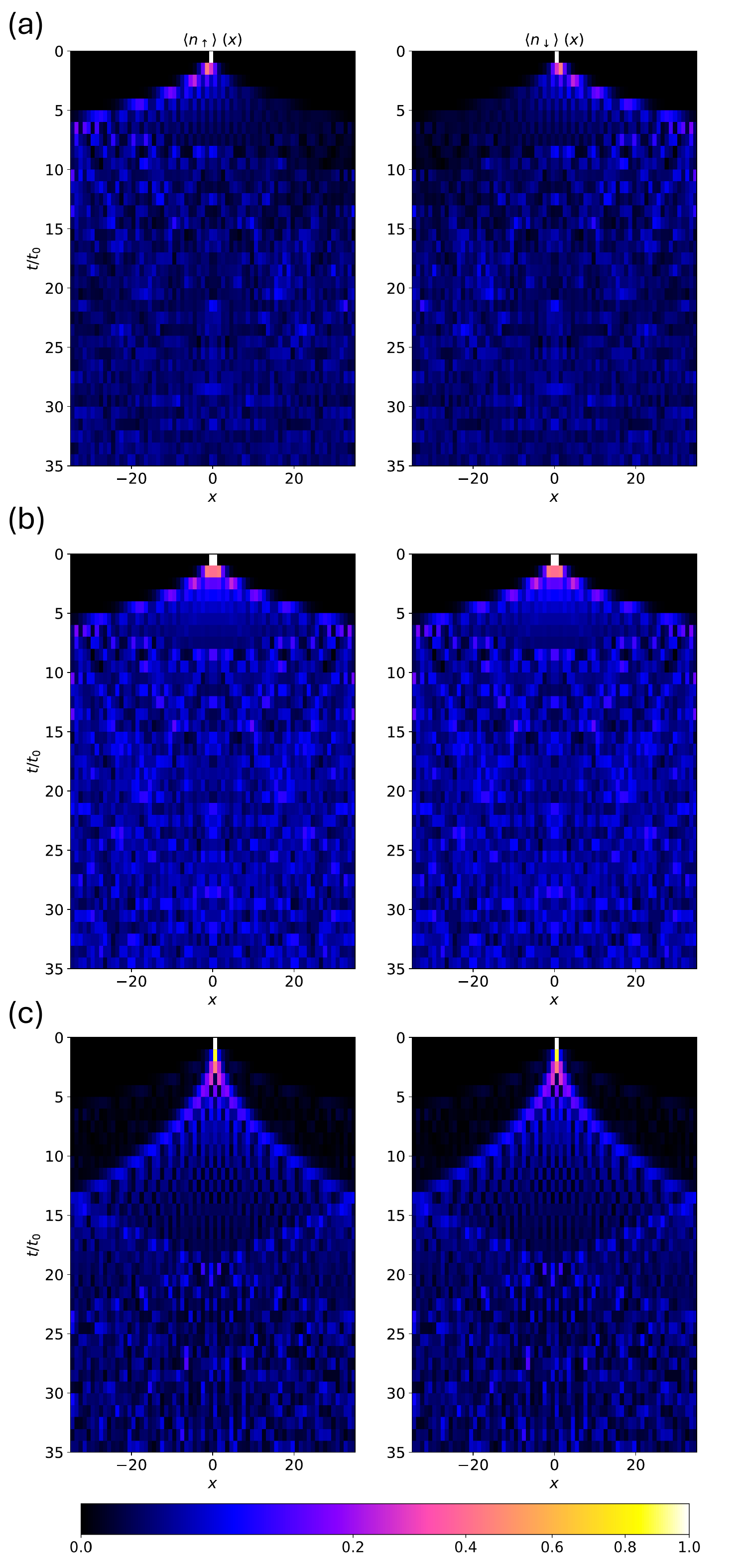}
    \caption{Occupation numbers $\langle n_\uparrow(x)\rangle$ (left column) and $\langle n_\downarrow(x) \rangle$ (right column) for the $\uparrow$ and $\downarrow$ distinguishable particles as a function of time with (a) separable, (b) entangled, and (c) doubly occupied initial states given in Eq. \eqref{eq:distinguishableProdInitalState}, Eq. \eqref{eq:distinguishableEntInitalState}, and Eq. \eqref{eq:distinguishableRightDoubleOccInitalState}, respectively. All walks were ran with a lattice of size $L=70$ and $U/J = 10$.
    }
    \label{fig:DistinguishableDynamicsPlots}
\end{figure}

Looking next at Fig. \ref{fig:DistinguishableESLRReduDoubleFig}(b) of the initially entangled state of Eq.~\eqref{eq:distinguishableEntInitalState}, we see that as we have set the initial state to already be maximally entangled between the two sides of the lattice, this unsurprisingly shows the maximal entanglement value from the first step and remaining constant throughout the walk. We find that this is not changed by the increase in the onsite repulsion, and the previously seen dips are eliminated outside of small fluctuations.

Lastly,  $ES_{LR,redu}$ with the doubly-occupied initial state  given in Eq. \eqref{eq:distinguishableRightDoubleOccInitalState} is shown in Fig, \ref{fig:DistinguishableESLRReduDoubleFig}(c). Here we see that it takes a similar form to the separable initial state of panel (a), with the reduction in the initial drop and fluctuations in the high onsite repulsion case that was seen for the separable initial state.

The left-right entanglement entropy  $ES_{LR,redu}$ for distinguishable particles, however, does not have a straightforward generalization to distinguishable particles since without the internal label for the particles, it is no longer possible to claim which particle is arranged in what patterns. In contrast, the left-right entanglement measure through coarse-grained states applies to distinguishable and indistinguishable cases on equal footing.

\section{Occupation-number plots of selected QW}
\label{sec:DynamicsPlots}
We show selected QW dynamics associated with each type of particles investigated (distinguishable and indistinguishable) for each of the initial state discussed in the main text. In Fig. \ref{fig:DistinguishableDynamicsPlots}, we show the time evolution of the $\uparrow$ and $\downarrow$ occupation numbers of two distinguishable walkers with $U/J=10$ on the left and right columns, respectively. Each row shows the results for the separable, entangled, and doubly occupied initial states given in Eq. \eqref{eq:distinguishableProdInitalState}, Eq. \eqref{eq:distinguishableEntInitalState}, and Eq. \eqref{eq:distinguishableRightDoubleOccInitalState}, respectively. We truncate the time evolution of each plot to $t/t_0=35$ as to not overcrowd the figures. We note that the case with doubly occupied initial state spreads at a slower rate, and doubly occupied states remain favorable throughout the dynamics. This is a consequence of energy conservation due to the unitary evolution and the relatively large initial interaction energy.

\begin{figure}[t]
    \centering
    \includegraphics[width=0.9\linewidth]{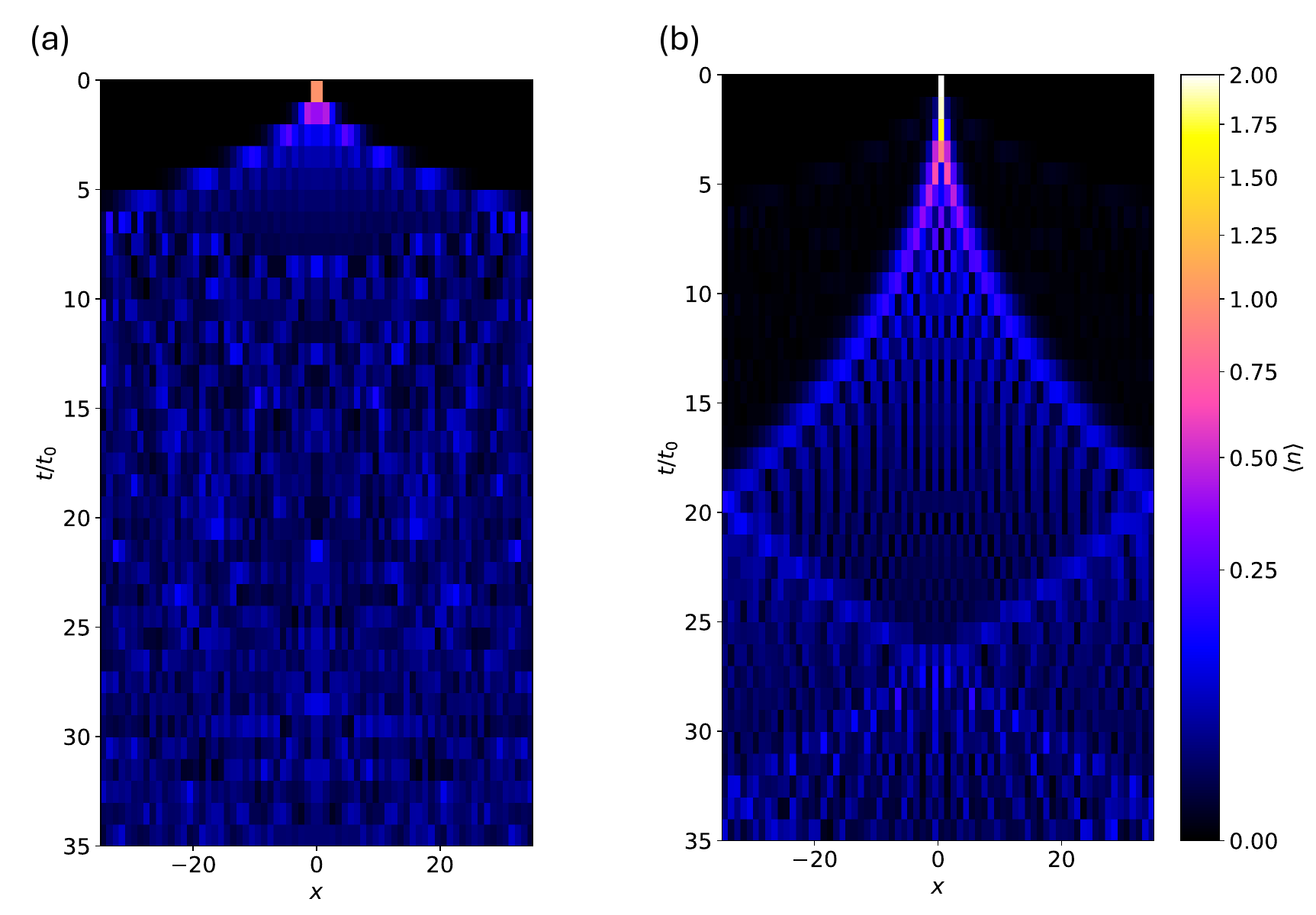}
    \caption{Occupation numbers $\langle n(x) \rangle$ for the two indistinguishable particles as a function of time for QWs with (a) adjacent singly, and (b) doubly occupied initial states given in Eq. \eqref{eq:indistinguishableAdjInitalState} and Eq. \eqref{eq:indistinguishableRightDoubleOccInitalState}, respectively. All walks were ran with a lattice size of $L=70$ and $U/J = 10$.
    }
\label{fig:IndistinguishableDynamicsPlots}
\end{figure}

Fig. \ref{fig:IndistinguishableDynamicsPlots} shows the time evolution of occupation numbers of two distinguishable walkers with $U/J=10$. Panels (a) and (b) show the results for the adjacent, singly occupied and doubly occupied initial states given in Eq. \eqref{eq:indistinguishableAdjInitalState}, and Eq. \eqref{eq:indistinguishableRightDoubleOccInitalState}, respectively. Once again, we truncate the time evolution of each plot to $t/t_0=35$ as to not overcrowd the figures. The doubly-occupied initial state shows a slower spreading Fig. \ref{fig:IndistinguishableDynamicsPlots} (b) compared to that of the singly occupied initial state. This because conservation of energy with higher interaction energy of the doubly occupied state makes it more challenging for the particles to move around in the lattice, which is consistent with the distinguishable cases shown in Fig.~\ref{fig:DistinguishableDynamicsPlots}.

%

\end{document}